\documentclass{rspublic}
\usepackage[colorlinks=true]{hyperref}
\usepackage{graphicx}
\usepackage{amsmath}
\usepackage{amssymb}
\usepackage{color}
\usepackage{bm}

\DeclareMathOperator{\sn}{sn}
\DeclareMathOperator{\cn}{cn}
\DeclareMathOperator{\dn}{dn}
\DeclareMathOperator{\am}{am}

\newcommand{\revision}[1]{{{#1}}}

\begin{document}
	
\title[Swarming, swirling and stasis in sequestered bristle-bots]{Swarming, swirling and stasis in sequestered bristle-bots}

\author[L. Giomi, N. Hawley-Weld and L. Mahadevan]{L. Giomi, N. Hawley-Weld and L. Mahadevan}

\affiliation{School of Engineering and Applied Sciences and Department of Physics, Harvard University, Pierce Hall 29 Oxford Street Cambridge, MA 02138, USA.\\[7pt]} 

\label{firstpage}		

\maketitle
	
\begin{abstract}{Swarming, collective behavior, robots}
The collective ability of organisms to move coherently in space and time  is ubiquitous in any group of autonomous agents that can move and sense each other and the environment. Here we investigate the origin of collective motion and its loss using macroscopic self-propelled Bristle-Bots, simple automata made from a toothbrush and powered by an onboard cell phone vibrator-motor, that can sense each other through shape-dependent local interactions, and can also sense the environment non-locally via the effects of confinement and substrate topography. We show that when Bristle-Bots are confined to a limited arena with a soft boundary, increasing the density drives a transition from a disordered and uncoordinated motion to organized collective motion either as a swirling cluster or a collective dynamical stasis. This transition is regulated by a single parameter, the relative magnitude of spinning and walking in a single automaton. We explain this using quantitative experiments and simulations that emphasize the role of the agent shape, environment, and confinement via boundaries. Our study shows how the behavioral repertoire of these physically interacting automatons controlled by one parameter translates into the mechanical intelligence of swarms.

\end{abstract}

\begin{center}
	\nointerlineskip
	\rule{0.9\textwidth}{0.7pt}
\end{center}

Collective behavior is ubiquitous among living organisms: it occurs in sub-cellular systems, bacteria, insects, fish, birds and in general in nearly any group of individuals endowed with the ability to move and sense (Miller 2010, Vicsek \& Zafeiris 2012). Recent studies of collective behavior have focused on the mechanism that triggers the switch from disordered to organized motion in a swarm (Vicsek \& Zafeiris 2012, Vicsek {\em et al}. 1995, Gregoire \& Chat\'e 2004, Ballerini {\em et al}. 2008, Leonard {\em et al}. 2012, Buhl {\em et al}. 2006), and its implications for artificially engineering these strategies in robotic systems (Mallouk \& Sen 2009, Rubenstein {\em et al}. 2011, Mellinger {\em et al}. 2010). For example, in social insects, such as the agarophilic desert locusts, the transformation from solitary to social behavior arises as a consequence of proximal tactile interactions that are density controlled (Buhl {\em et al}. 2006). Experiments on the claustrophilic termites, {\it Macrotermes michaelseni} which are used to living in confined spaces, have demonstrated the existence of a variety of collective behaviors such as coordinated circulation and arrest or stasis in a closed confined geometry. These different behaviors may be triggered by varying the density of the colony and disturbing it through external stimuli (Turner 2011). Understanding how these biological behaviors arise from a mechanistic perspective has been difficult given our primitive experimental abilities to probe the neuro-ethology of these complex creatures. Theoretical attempts to understand these behaviors use putative models of interactions between organisms as a function of their density in periodic domains (Vicsek \& Zafeiris 2012), while a practical approach circumvents the question of mechanism and implements workable strategies to actively direct the collective dynamics of ensembles of agents (Rubenstein {\em et al}. 2011, Mellinger {\em et al}. 2010) using feedback control  in individual agents (Braitenberg 1984). These approaches clarify the common bases at the heart of all swarming behaviors: the ability of an agent to move, the ability to sense others and the environment, and the ability to respond to both of these kinds of stimuli.

\begin{figure}[t]
\centering
\includegraphics[width=1\textwidth]{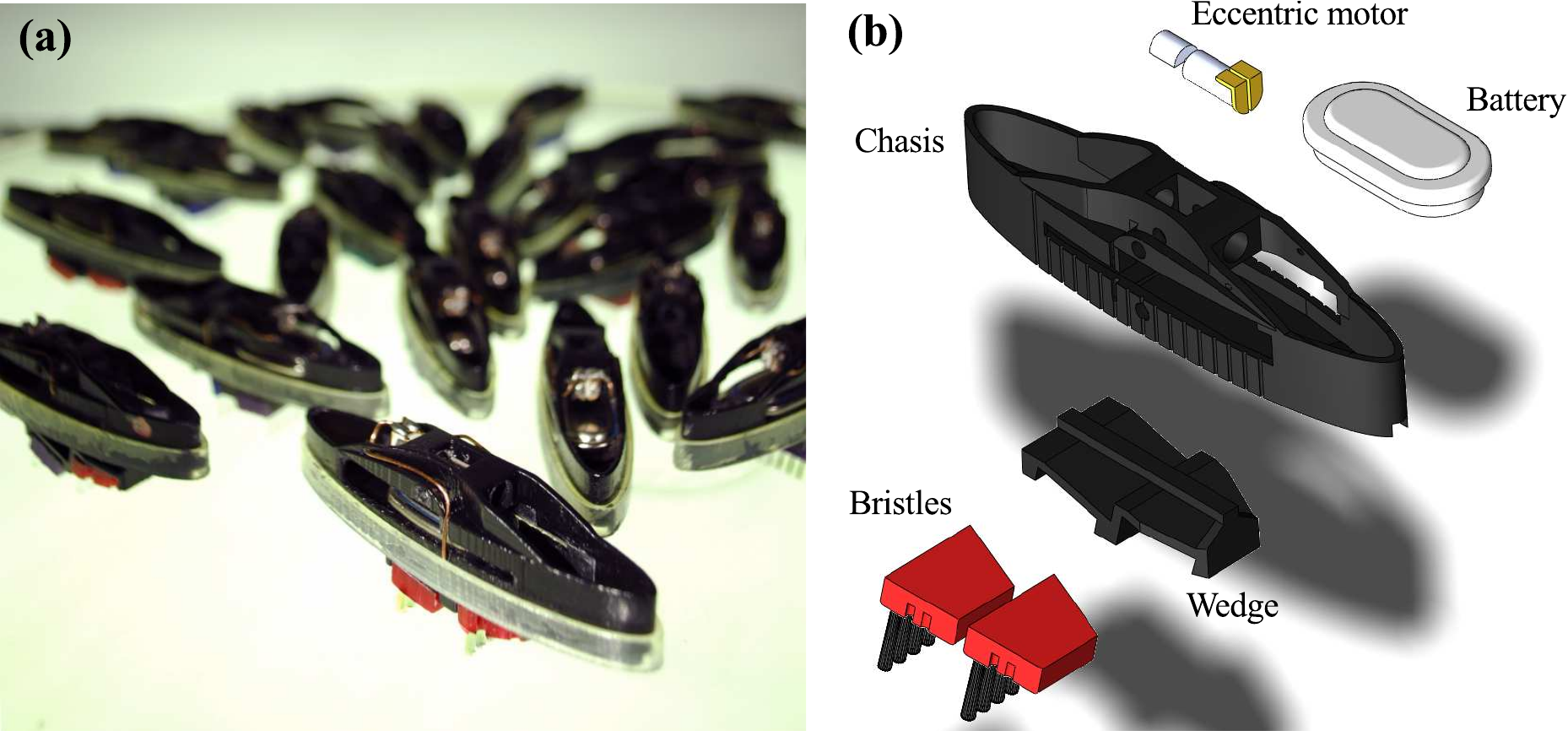}	
\caption{\label{fig:design}(a) A collection of the BBots used in the experiment. (b) Schematic of an individual BBot. A plastic chassis is connected to a pair of toothbrushes via a slanted wedge. An eccentric motor is positioned on the top side of the device and is powered by a VARTA rechargeable button-cell battery.}
\end{figure}

Here, we probe the transition from random swarming to collective motion and its loss using a minimal system composed of self-propelled automatons that can sense each other mechanically through contact and interact both with an environment of varying topography and with boundaries. Our setting is macroscopic, controllable and especially suitable to investigate the role of the environment in selecting and tuning the collective behavior of the group. \revision {Unlike experiments on vibrated particles (e.g. Narayan {\em et al}. 2007, Deseigne {\em et al}. 2010), where all particles are simultaneously driven using the same source, our agents are autonomous and self-propelled, with velocities that are independent, and yet show collective behavior even in a small group of individuals in the presence of confinement.}

\section{Motion of an individual BBot}

Our experiments were carried out using a custom-fabricated swarm of Bristle-Bots~\footnote{See also: \url{http://www.evilmadscientist.com/article.php/bristlebot}
} (BBots) (Murphy 2009, Bobadilla {\em et al}. 2011), simple self-propelled  automata with similarities both to natural mechanical ratchets (Kuli\'c {\em et al}. 2009) and their artificial analogs (Mahadevan {\em et al}. 2004). Our system has three controllable features: (i)  a tunable ratio of linear  speed and rate of turning for individual agents; (ii) a collective ability to exert aligning forces and torques on each other by means of shape dependent contact interactions; (iii) confinement induced by soft or hard boundaries. The design of our BBots (Fig. \ref{fig:design}) is optimized to be small, light, stable and modular. An elliptical plastic chassis (major axis 7.92 cm, minor axis 1.85 cm) serves as a container for a 1.2V VARTA rechargeable battery which can slide inside the chassis to adjust the position of the center of mass and thus change the relative ratio of translational and rotational speed. The battery is connected to a motor (commonly used in cell phones) housed on the top side of the chassis, with a mass of 0.5 g and an eccentricity of 0.8 mm, designed to rotate at 150 rounds per second. Two rows of nylon bristles, obtained from a commercial toothbrush, form the legs of our BBots. The bristles are cut to 5 mm length to prevent tipping without compromising their flexibility and are attached to the chassis via a removable wedge. This allows us to control the inclination of the bristles relative to the chassis.  The total mass of the object is 15.5 g. 

\begin{figure}[t]
\centering
\includegraphics[width=1\textwidth]{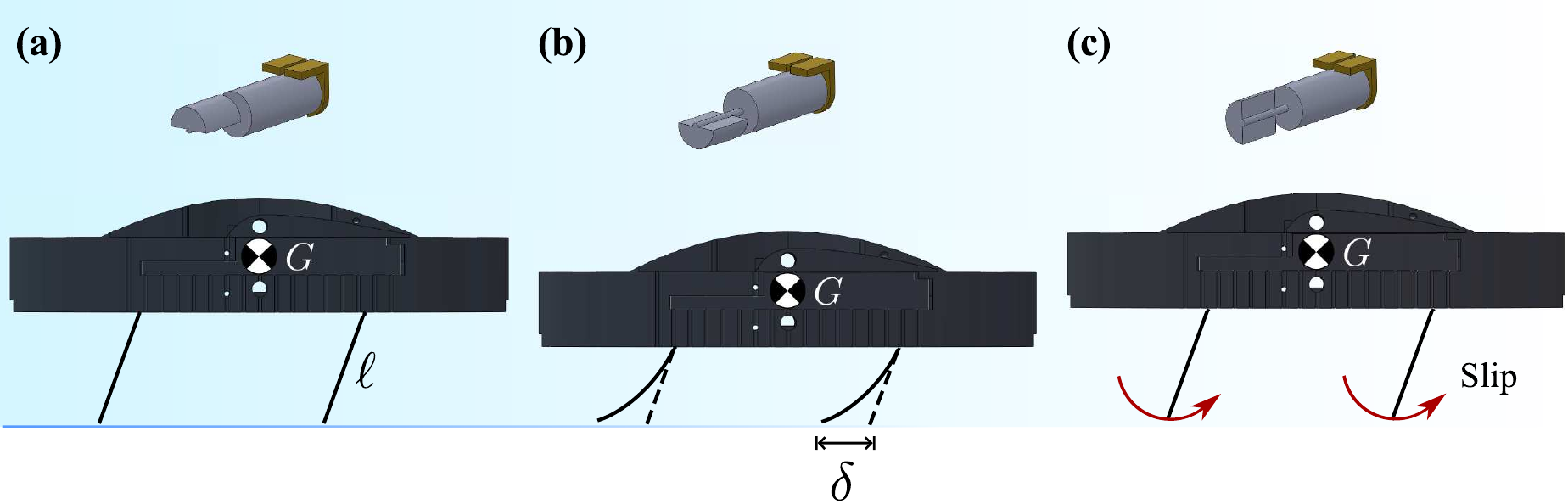}	
\caption{\label{fig:locomotion}Principle of motion of a single BBot. The bristles act as legs that are periodically flexed under the action of the eccentric motor. Over a cycle, the bending and unbending of the tilted bristles causes them to slip, resulting in forward motion.}
\end{figure}

\begin{figure}
\centering
\includegraphics[width=0.8\textwidth]{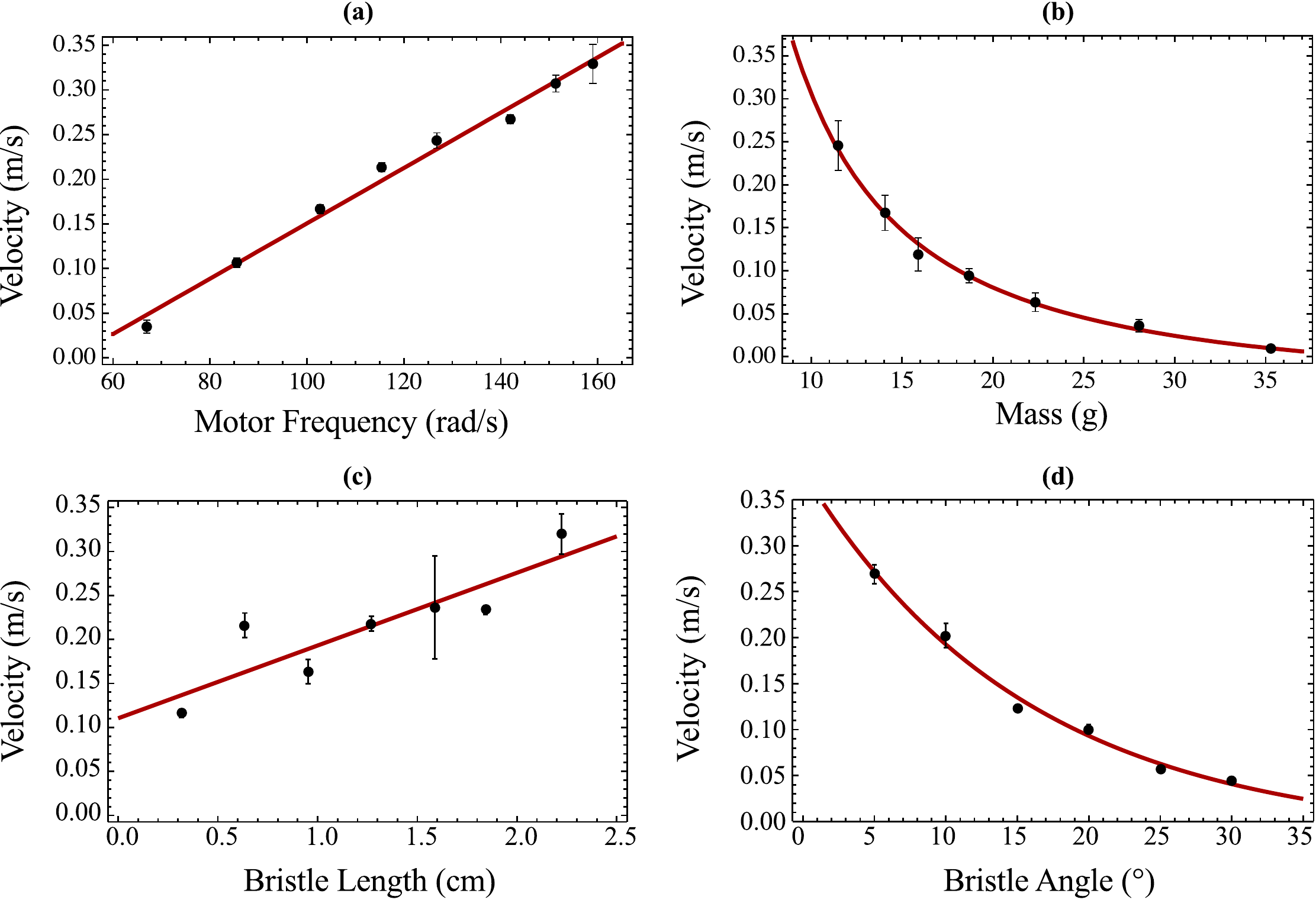}	
\caption{\label{fig:single-bot}Performance of a Bbot, i.e. its velocity when confined to a channel, as a function of (a) motor frequency, (b) total mass, (c) bristle length and (c) inclination with respect to the vertical.}
\end{figure}

BBots move when the eccentrically loaded motor drives the legs of the machine, the bristles, which flex periodically. The bending of the tilted bristles on the substrate causes them to move more easily in the forward direction relative to the rear, leading to a rectification of the periodic driving and thus directed movement. Over each period of rotation of the eccentric motor the sequence shown in Fig. \ref{fig:locomotion} is followed (see also \ref{sec:appendix-a} and Supplementary Movie 1): 1) the bristles are loaded by a force $F=Mg+mr\omega^{2}$ ($Mg$ weight of the BBot, $m$ eccentric mass, $r$ lever arm, $\omega$ angular frequency of the motor); 2) as the eccentric mass rotates, the load on the bristles decreases, causing the bristles to recoil; 3) the bristles slip forward on the underlying substrate, producing a net displacement of the object. To quantify the motion of an individual BBot, we analyzed the shape of a row of bristles treated as a single elastic beam subject to a periodic tip load as well as a frictional force in the horizontal direction (see \ref{sec:appendix-a}) and showed how the linear velocity of a BBot and its turning rate depend on the design parameters of the system.  

In Fig. \ref{fig:single-bot}, we show how the motor speed, bristle position, length and angle and the system mass leads to changes in the speed of an individual BBot confined to a narrow channel to prevent lateral drift. We see that the bristle inclination and length have a strong effect on BBot locomotion; increasing the angle $\alpha$ of the bristle with respect to the vertical direction causes the BBots to slow down substantially when $\alpha$ is varied from $5^{\circ}$ to $30^{\circ}$. The length of the bristles affects the motion of a BBot in two ways: longer bristles cause the center of mass to be displaced further in each step, leading to a linear increase in the velocity (Fig. 1c-3), while short stiff bristles lead to a noisier dynamics associated with rebounds and jumps driven by the eccentric forcing. Furthermore, because long bristles cause the BBots to spend a longer time in contact with the substrate (where the transverse component of the eccentric force is balanced by friction), they move primarily along a straight line, while BBots equipped with short bristles are prone to move in a circle. This sensitive dependence on the bristle parameters allows us to tune the locomotion of individual subunits and study its role on the collective behavior of the community. In particular, by choosing 5 mm bristles and varying $\alpha$ from $0^{\circ}$ (upright bristles) to $20^{\circ}$, we obtained two distinct types of individuals: 1) spinners, which are BBots with $\alpha=0^{\circ}$ and $5^{\circ}$ that tend to spin clockwise with an angular velocity of up to 30 rad/s while moving slowly; 2) walkers, which are BBots with $\alpha \ge 10^{\circ}$ that move in a straight or weakly curved orbit (Supplementary Movie 2).

\begin{figure}[t]
\centering
\includegraphics[width=0.7\textwidth]{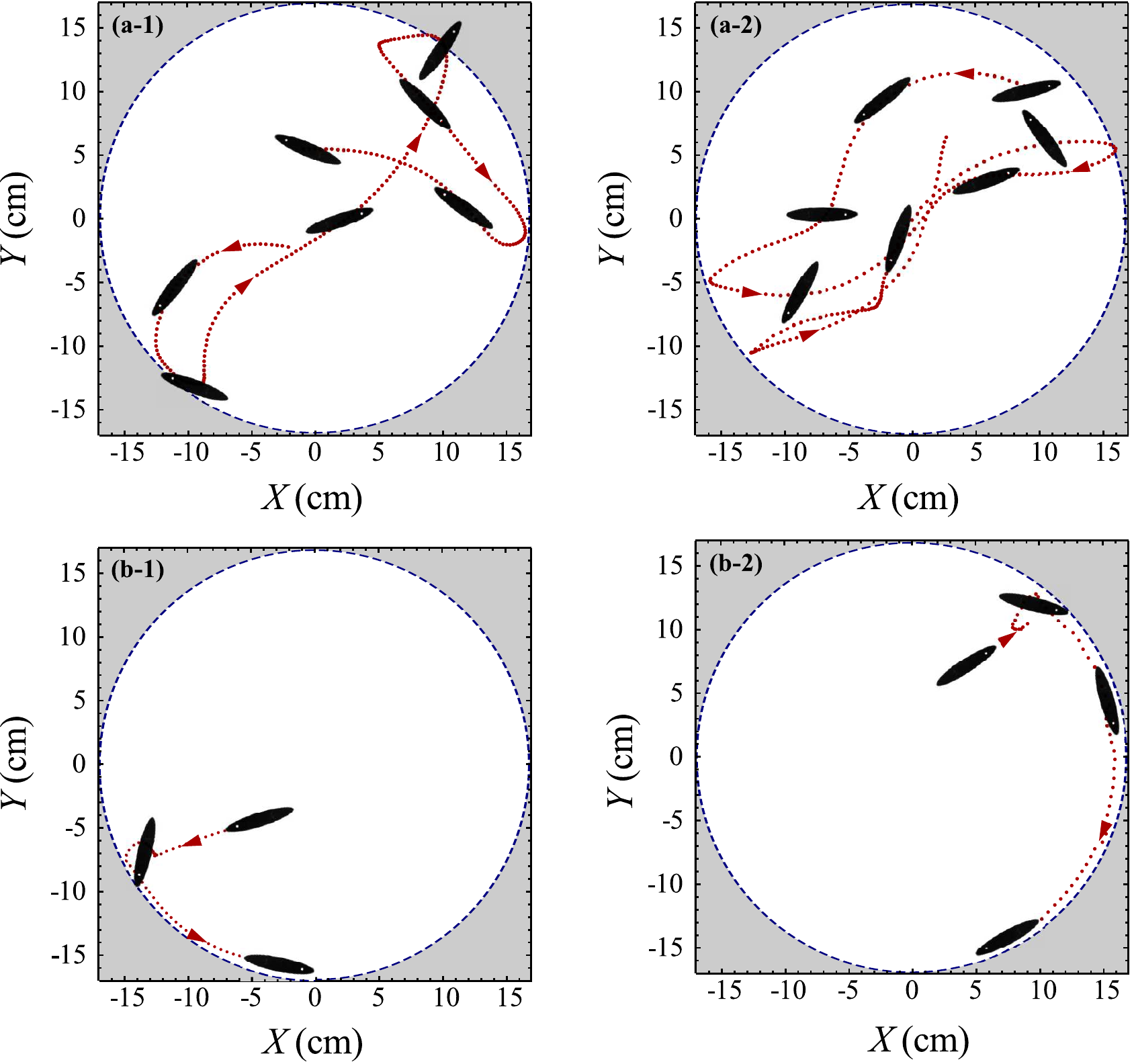}	
\caption{\label{fig:trajectory}{Experiments showing the interaction of BBot with a soft and hard boundary. (a-1) A spinner with $\alpha=5^{\circ}$ and (a-2)  a walker with $\alpha=10^{\circ}$  in a soft-boundary arena. (b-1) A spinner with $\alpha=5^{\circ}$  and (b-2) a walker with $\alpha=10^{\circ}$  in a hard-boundary arena. For a soft boundary,  a consequence of the shallow bowl-shaped curvature is that the BBot is reflected toward the interior. For a hard boundary, the BBot gets aligned with the edge of the arena and moves along it.}}
\end{figure}

\section{The effect of boundaries and topography}

In most prior studies of collective behavior, the boundary is not considered; indeed theoretical studies routinely treat only the case with periodic boundaries, while the few experimental studies that exist aim to minimize the role of boundaries. In our study, the boundaries play a most important role as we now discuss. Our arena consists of a circular plate of 44 cm in diameter, with a single BBot taking up approximatively 0.8\% of the total available area. 

\subsection{Soft boundaries}

We first consider the interaction of a BBot with the boundary that causes it to be reflected back from the edge, into the middle, using an arena with a gentle upward sloping edge, fabricated by oven-forming an acrylic disc over a frisbee-shaped aluminum mold. Here, we see that surface topography plays a role normally reserved for the boundary by influencing the motion of a BBot via environmental changes. With this {\em soft boundary} setup, our BBots either turn back into the middle (behavior typical of spinners, i.e. $\alpha=0^{\circ},\,5^{\circ}$) or they oscillate back and forth in a periodic motion that causes them to remain in the neighborhood of a particular location at the boundary (behavior typical of walkers, i.e. $\alpha=10^{\circ},\,15^{\circ}$) (Fig. \ref{fig:trajectory}a). This pendulum-like effect follows from the fact that the walker's path is never perfectly radial, so that as a BBot climbs the edge it also turns sideways. On the steepening gradient near the edge, the BBot typically slips backwards, as it rotates by about $30^{\circ}$, and Sisyphus-like, tries to climb up the edge again only to be kicked back to where it started. These oscillations may be repeated a few times for an individual Bbot  before it eventually moves back into the center of the arena, and then onto another part of the edge where the same phenomena is repeated. Strong walkers with $\alpha=20^{\circ}$ do not experience the oscillatory motion at the boundary because their forward propulsion dominates the role of sideways spinning motion and tends to align the BBots to be normal to the edge independent of how they initially approach the boundary.

\subsection{Hard boundaries}

To see what happens when we change the environment in which the Bbots operate, we replaced the boundary of the arena with a gently curved edge with a flat circular disc of the same 44 cm diameter, bounded by a thick (vertical) strip of acetate  $\sim 4$ cm high that is held firmly in place by a ring of thick translucent tubing.  The most salient feature of this {\em hard boundary} system is that the boundaries are not reflective, so that a BBot that hits the edge will begin to circulate in a particular direction around the arena, traveling always parallel to the edge (Fig. \ref{fig:trajectory}b). We observe stable motion in both a clockwise and counterclockwise direction, the determining factor being the angle of initial contact with the wall. 

\section{Collective behavior of BBots - Experiments}

\begin{figure}[h]
\centering
\includegraphics[width=0.9\textwidth]{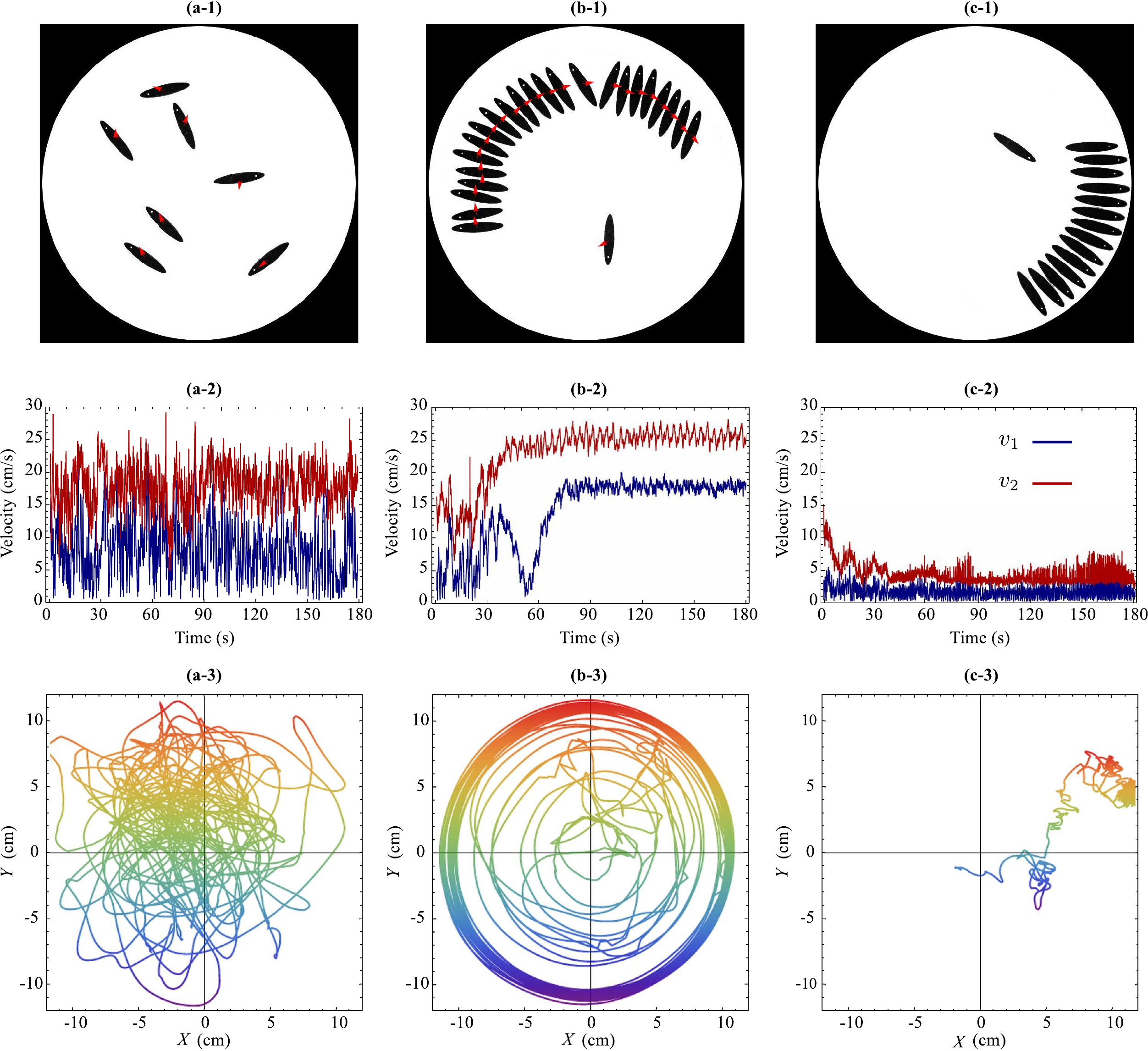}	
\caption{\label{fig:collective-behavior-ex}Experimental realization of collective behavior of Bbots. The columns summarize the three behaviors observed in the experiments with BBots: (a) disordered (random) motion of spinners at low density; (b) swirling motion of spinners at high density, and (c) stasis of walkers at high density. (a1-a3) Experiments in the random phase; (a1) instantaneous position of 7 spinners with $\alpha=5^{\circ}$, (a2) the mean velocity $v_{1}$ and the mean speed $v_{2}$ of the Bbots, (a3) the trajectory of the center of mass of the Bbots in physical space, showing random motion. (b1-b3) Experiments in the swirling phase; (b1) Instantaneous position of 24 spinners with $\alpha=5^{\circ}$, (b2) $v_{1}$ and $v_{2}$ showing a non-zero value, (b3) the trajectory of the center of mass of the Bbots in physical space, showing the signature of the coordinated swirling. (c1-c3) Experiments in the stasis phase; (c1) Instantaneous position of 15 walkers with $\alpha=10^{\circ}$, (c2) $v_{1}$ and $v_{2}$ of the Bbots, (c3) the trajectory of the center of mass of the Bbots in physical space, showing no motion, i.e. stasis.}
\end{figure}

To study the collective dynamics of the BBots,  we use a transparent plate that is backlit with a set of neon lamps and allows us to track the BBots with a digital camera at 40 fps. The resulting movies were processed with tracking software to compute the position, orientation and the translational and rotational velocity of each BBot and thus quantify their individual behavior. This allows us to calculate the following two order parameters to characterize the collective behavior of the putative swarm: 
\begin{equation}
v_{1}(t) = \frac{1}{N}\left |\sum_{i=1}^{N}\bm{v}_{i}(t)\right|\;,\qquad
v_{2}(t) = \frac{1}{N}\sum_{i=1}^{N} \left|\bm{v}_{i}(t)\right|\;.
\end{equation}
where $N$ is the total number of BBots and $\bm{v}_{i}(t)$ the velocity of the $i-$th BBot at time $t$. We see that $v_{1}$ is the average velocity of the BBots, while $v_{2}$ is their average speed. When they move in a disordered fashion, $v_{1}\approx 0$ (becoming exact in the infinite particle limit) and $v_{2}>0$; BBots moving coherently in space have both $v_{1}\ne 0$ and $v_{2}\ne 0$, while if a cluster of BBots is dynamically arrested, $v_{1} \approx v_{2} \approx 0$. 

In our experiments we used a paperboard template that initially arrested the motion of the BBots. When this was removed, the BBots moved and eventually reached a statistical steady state (Supplementary Movie 3). Alternatively, we progressively increased the BBot population from 2 to 16, adding a new one every 30 seconds and then removing them one at a time, to measure the hysteresis in the transition between states of collective behavior. 

In Fig. \ref{fig:collective-behavior-ex}a,b, we show the results of our experiments on the collective motion of spinners ($N = 7$, 24 with $\alpha=5^{\circ}$) moving in an arena with a soft boundary for 3 minutes. Spinners spin rapidly and collide frequently and strongly with each other (Supplementary Movie 3); when $N<10$ (corresponding to 8\% area coverage) their motion  is disordered, with $v_{1}<v_{2}$ as seen in Fig. \ref{fig:collective-behavior-ex}a-2 and their center of mass moves aperiodically as shown in Fig. \ref{fig:collective-behavior-ex}a-3. When $N>10$ the spinners aggregate at the edge of the arena   while aligning themselves at an angle to the boundary, and start swirling collectively clockwise (the direction of spinning for individual BBots) coherently along the edge, as show in Fig. 2b-2. In this case, the order parameter $v_{1}$ increases and saturates once the swirling cluster is formed (Fig. \ref{fig:collective-behavior-ex}b-2). 

\begin{figure}[t]
\centering
\includegraphics[width=0.9\textwidth]{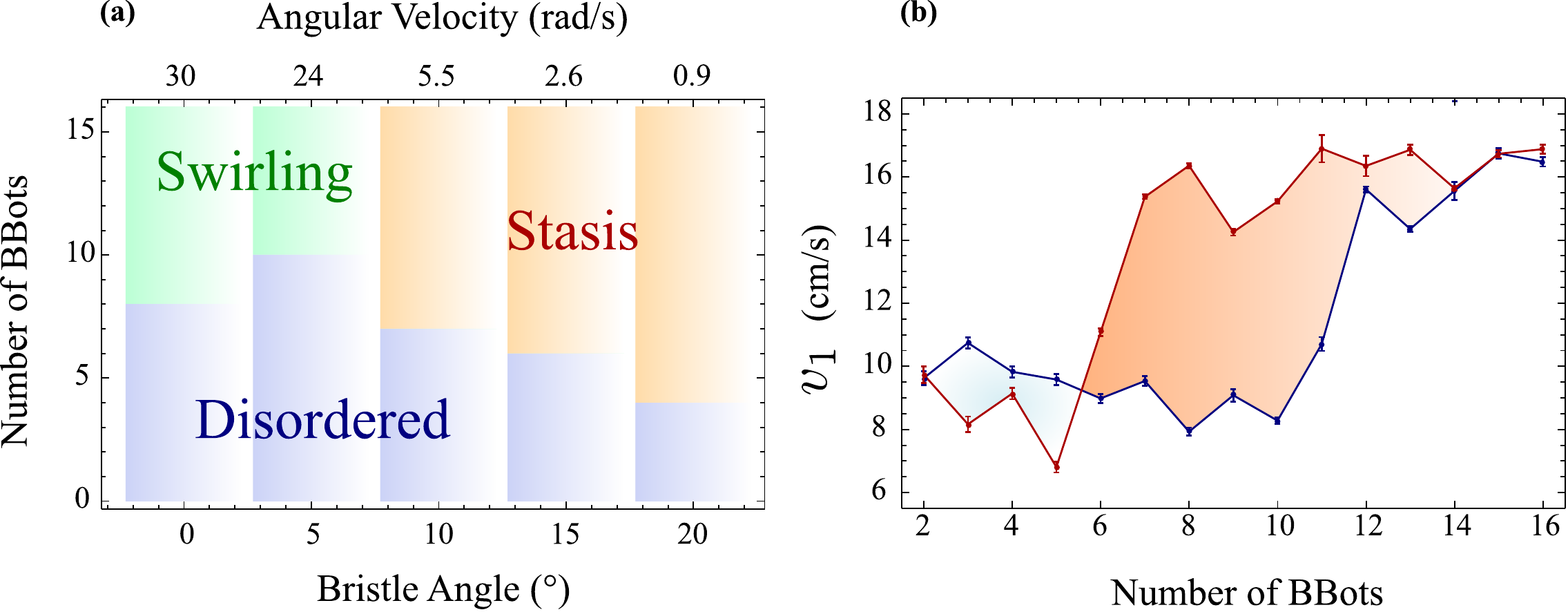}	
\caption{\label{fig:phase-diagram}Phase diagram and hysteresis in collective behavior. (a) A phase diagram summarizes the three types of collective behavior observed as a function of the number of BBots (which is equivalent to their density) and the bristle angle (which is proportional to the inverse of the velocity) confined to an arena with a soft boundary. (b) Hysteresis diagram obtained by progressively increasing and decreasing the number of spinners in the range $2\le N \le 16$. In the forward portion of the curve (blue) the population transitions from disordered to swirling motion for $N>10$. After the onset of collective motion, newly added BBots are eventually collected by the swirling cluster. When BBots are withdrawn from the trailing edge of the swirling cluster the behavior switches from coordinated to disordered only when the population is below 6 BBots.}
\end{figure}

Walkers at low density have a different behavior than spinners; they move to the edge, stay for a while before they turn around randomly, eventually reaching an approximately antipodal point where this behavior is repeated. As the number of walkers is increased, they form ephemeral clusters along the edge (Supplementary Movie 4) that eventually break up. However, when $N>8$, clusters of BBots oriented perpendicular to the edge form and remain stable, as shown in Fig. \ref{fig:collective-behavior-ex}c ($N=15$ with $\alpha=10^{\circ}$). This corresponds to the order parameters $v_{1}\approx v_{2}\approx 0$, and the center of mass is essentially stationary (Fig. \ref{fig:collective-behavior-ex}c-2 and \ref{fig:collective-behavior-ex}c-3). In Fig. \ref{fig:phase-diagram}, we show a phase diagram that summarizes the collective behavior of BBots confined to an arena with a soft boundary, showing disordered motion, swirling and stasis, and highlights the hysteretic nature of the transitions between states. For example, once a swirling cluster of spinners has formed, it remains stable even when BBots are withdrawn from the cluster until $N<6$.

\begin{figure}[t]
\centering
\includegraphics[width=0.8\textwidth]{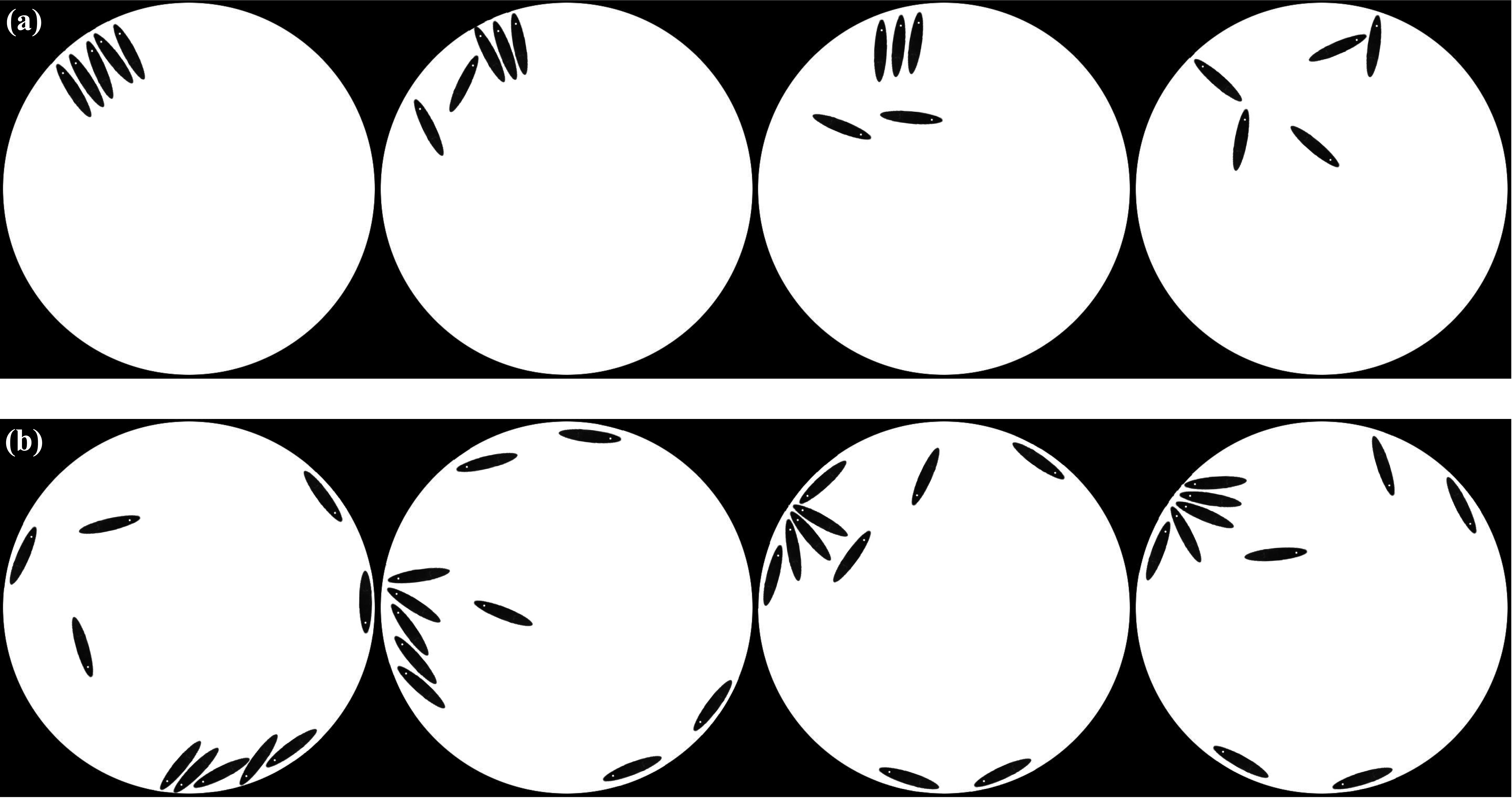}	
\caption{\label{fig:cluster}An experimental example of cluster assembly/disassembly in the presence of soft and rigid boundary. (a)  Five walkers with $\alpha=10^{\circ}$ initially gather at the soft boundary. The cluster, however, starts to oscillate and eventually disassembles. (b) The same walkers in an arena with a hard boundary corresponding to a vertical wall jam to form a half-aster pattern which nucleates and grows in size until all the BBots in the system have been collected by the jammed cluster.}
\end{figure}

To understand how confinement and topography lead to these behaviors, we first use BBots with an intermediate bristle inclination to observe the assembly and disassembly of clusters at a soft boundary; for example, BBots with  $\alpha = 10^{\circ}$ resist the sideways motion necessary for swirling but they do not get trapped at the edge as easily as BBots with larger $\alpha$. The result is that clusters can form orthogonal to the circular boundary, but if the cluster is too small in  number it will eventually disassemble due to the growth of coordinated oscillations of the entire cluster (Fig. \ref{fig:cluster}a). Indeed clusters of four or five bots remain stable for over a minute before disassembling. The stasis or jamming region in our phase diagram describes the formation of clusters at even higher densities when they become stable over very long times.

In contrast, collective behavior in the presence of a hard-boundary leads to contact with the vertical wall and aligns the BBots along the boundary thus limiting their motion and reducing the interactions between BBots (Fig. \ref{fig:trajectory}b). BBots sliding along the boundary of the arena eventually form groups due to the small variations in the velocity of individuals. However, as the number of Bbots in a group increases, it becomes less stable and can abruptly self-arrest. These arrested states can take the shape of a half-aster as shown in Fig. \ref{fig:cluster}b when the arena is bounded by a vertical, rigid boundary, in contrast with the orthogonally oriented jammed structures formed in the soft boundary arena.

\section{Collective behavior of BBots - Theory}

\begin{figure}[t]
\centering
\includegraphics[width=0.4\textwidth]{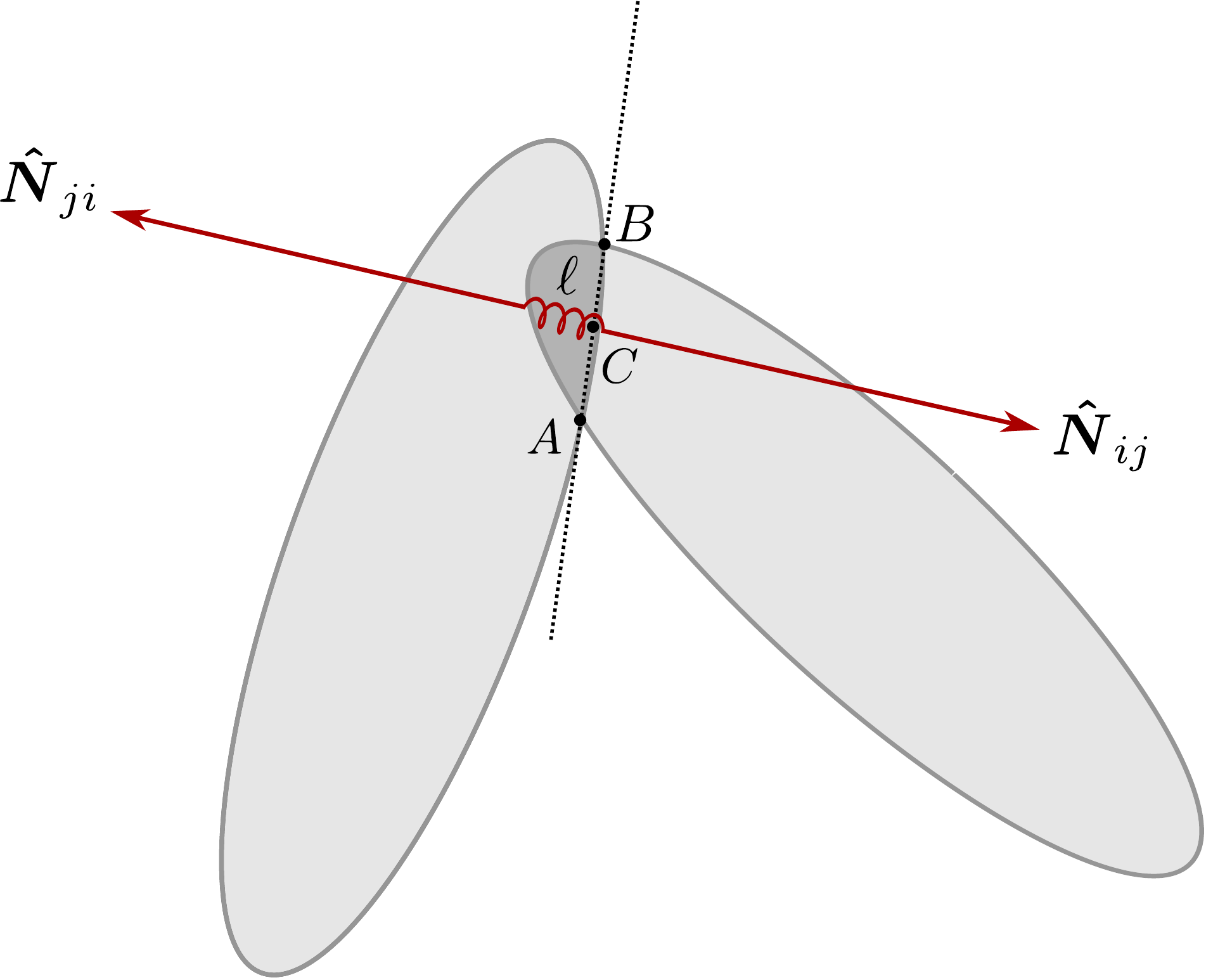}
\caption{\label{fig:spp-forces}Schematic representation of the forces between overlapping   ellipses. The forces are applied along the direction $\bm{N}_{ij}$ perpendicular to the line passing through the intersection points $A$ and $B$ of the two particles, at the mid point $C$. The magnitude of the force is controlled by the spring-length $\ell$ obtained by intersecting the line $\bm{N}_{ij}$ with the perimeter of the region where the two ellipses overlap (shaded in the figure). The overlap between the particles is exaggerated in the figure; in the simulations it is very small, so that the direction $\bm{N}_{ij}$ approximates the common normal direction of two convex objects touching at one point.}	
\end{figure}

The nature of the collective motion and stasis in our system of confined agents relies on the ability of the BBots to march in the direction of their major axis, and rotate and align with each other and with the boundary. In order to understand these effects quantitatively, we use simulations of self-propelled particles (SPP) consisting of two-dimensional ellipses whose center of mass position $\bm{r}_{i}$ and orientation $\theta_{i}$ are governed by the following dynamical system:
\begin{subequations}\label{eq:model}
\begin{align}
\frac{d\bm{r}_{i}}{dt} &= v_{0}\bm{n}_{i}+k_{1}\sum_{j=1}^{N_{i}}\bm{F}_{ij}\\[7pt]
\frac{d\theta_{i}}{dt} &= \omega+\zeta_{i}+k_{2}\sum_{j=1}^{N_{i}}M_{ij}
\end{align}
\end{subequations}
The first equation describes the over-damped motion of individual ellipses with velocity $v_{0}$ along their major axis $\bm{n}_{i}=(\cos\theta_{i},\sin\theta_{i})$, where  $\bm{F}_{ij}$ is the repulsive elastic force between the $i-$th ellipse and its $N_{i}$ neighbors, these being defined as the set of all ellipses that overlap with the $i-$th. This force between the $i$-th particles and its $N_{i}$ neighbors is given by
\begin{equation}
\bm{F}_{ij} = \ell\,\bm{\hat{N}}_{ij}\,,
\end{equation}
with $\ell$ a virtual {\em spring length}, which for the ellipses is calculated from the intersections between the two overlapping ellipses as illustrated in Fig. \ref{fig:spp-forces}.

The second equation implies that the major axis of each ellipse rotates counterclockwise with frequency $\omega$ and can align with its neighbors as a consequence of the physical torque due to the contact with the neighbors and is given by:
\begin{equation}
M_{ij} = (\bm{d}_{ij}\times\bm{F}_{ij})\cdot\bm{\hat{z}}\,,
\end{equation}
where $\bm{d}_{ij}$ is the lever arm of the force $\bm{F}_{ij}$ exerted by the $j-th$ neighbor on the $i-th$ ellipse and $\bm{\hat{z}}$ is the unit vector in the $z$-direction. The constant $k_{2}$ measures the strength of this aligning interaction, while $\zeta_{i}$ is a delta-correlated random variable in the interval $[-\zeta,\zeta]$ and represents the  noise associated with all the non-deterministic factors that affect our system. 

The ellipses are confined to a circular arena of radius $R$ and subject to a non-local exponentially decaying torque exerted by the boundary that reorients them toward the interior, which reflects the torque produced by the curvature of the experimental arena along the edge. The interaction between the particles and the boundary takes place through a virtual linear spring acting at the center of mass: 
\begin{equation}
\bm{F}_{i}^{\rm boundary} = -k_{1}(|\bm{r}_{i}|-R)\,\bm{\hat{r}}_{i}\,, \qquad r>R
\end{equation}
and a long range torque of the form:
\begin{equation}\label{eq:boundary-torque}
M_{i}^{\rm boundary} = -k_{3}\sin[\arctan(y_{i}/x_{i})-\theta_{i}]\exp\left(\frac{|\bm{r}_{i}|-R}{\xi}\right)\,.
\end{equation}
where $\xi$ is a constant length that can be used to tune the range of the interaction. When a particle is in proximity of the boundary [i.e. $(|\bm{r}_{i}|-R)/\xi \approx 1$], this boundary torque has the effect of rotating the particle toward the interior. The non-locality of the boundary torque $M_{i}^{\rm boundary}$ in \eqref{eq:boundary-torque} mimics the distributed gravitational torque produced by the curvature of the dish in which the particles move and  appears to have a significant role for the clustering of the particles at the boundary. Compared with its local-analog (i.e. a torque of the same form that acts only when $|\bm{r}_{i}|>R$), the torque \eqref{eq:boundary-torque} has the effect of producing more densely packed clusters of particles along the boundary and thus fundamentally changes the collective behavior of the Bbots. Being curved, the boundary of the arena has the effect of getting the particles to form densely packed clusters. Extending the range of the particle-boundary interaction is equivalent to increasing the curvature and thus accentuates the focusing effect. This is also the simplest situation where we see how the physical environment can control the behavior of these autonomous agents. 

Our system differs fundamentally from those studied in the past by accounting correctly for orientational effects using torque balance rather than an ad-hoc alignment term, while also exploring the role of non-local interactions using topography and finite size boundaries. The dynamical system \eqref{eq:model} is characterized by four dimensionless parameters: the scaled density $\phi=R^{2}/ab$ (where $a$ and $b$ are the minor and major semi-axes of the ellipses), the spinning to walking ratio $\omega a/v_{0}$, the orienting parameter $k_{2}a^{2}/k_{1}$ and the scaled noise parameter $\zeta a/v_{0}$. In our experiments, the relevant experimental variables are the scaled density and the spinning ratio, since the orienting parameter and the noise are intrinsic to the shape of the agents and the motor characteristic. 

\begin{figure}[t]
\centering
\includegraphics[width=0.9\textwidth]{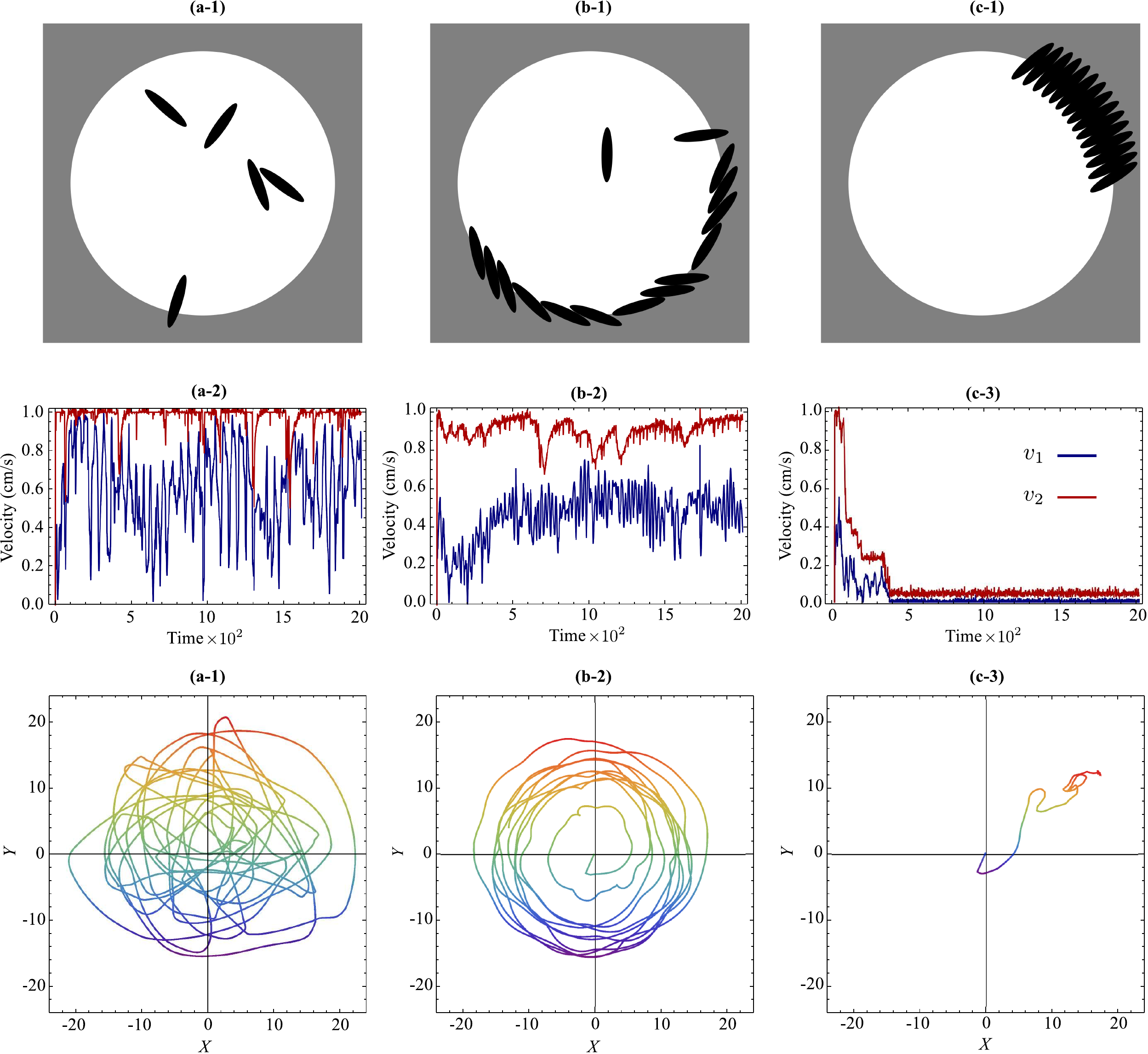}
\caption{\label{fig:collective-behavior-th}Numerical simulations of collective behavior of Bbots. As in Fig. \ref{fig:collective-behavior-ex} the columns summarize the three behaviors observed in the experiments with BBots: (a) disordered (random) motion of spinners at low density; (b) swirling motion of spinners at high density, and (c) stasis of walkers at high density. (a1-a3) Numerical solution of Eqs. \eqref{eq:model} in the random phase, with $\omega a/v_{0}=0.05$; (a1) Instantaneous position of 5 spinners with $\alpha=5^{\circ}$, (a2) the mean velocity $v_{1}$ and the mean speed $v_{2}$ of the Bbots, (a3) the trajectory of the center of mass of the Bbots in physical space, showing random motion. (b1-b3) Numerical solution of Eqs. \eqref{eq:model} in the random phase, with $\omega a/v_{0}=0.03$; (b1) Instantaneous position of 15 spinners with $\alpha=5^{\circ}$, (b2) $v_{1}$ and $v_{2}$ showing a non-zero value, (b3) the trajectory of the center of mass of the Bbots in physical space, showing the signature of the coordinated swirling. (c1-c3) Numerical solution of Eqs. \eqref{eq:model} in the random phase, with $\omega a/v_{0}=0$; (c1) Instantaneous position of 15 walkers with (c2) $v_{1}$ and $v_{2}$ of the Bbots, (c3) the trajectory of the center of mass of the Bbots in physical space, showing no motion, i.e. stasis. For all simulations, we chose the ellipse aspect ratio to be 5, while the other parameters are $k_{1}a/v_{0}=10$, $k_{2}a^{2}/v_{0}=1$ and $\zeta a/v_{0}=2\pi$.  The equations are integrated via a four-step Runge-Kutta algorithm with time step $\Delta t=0.001$.}	
\end{figure}

\begin{figure}[t]
\centering
\includegraphics[width=0.9\textwidth]{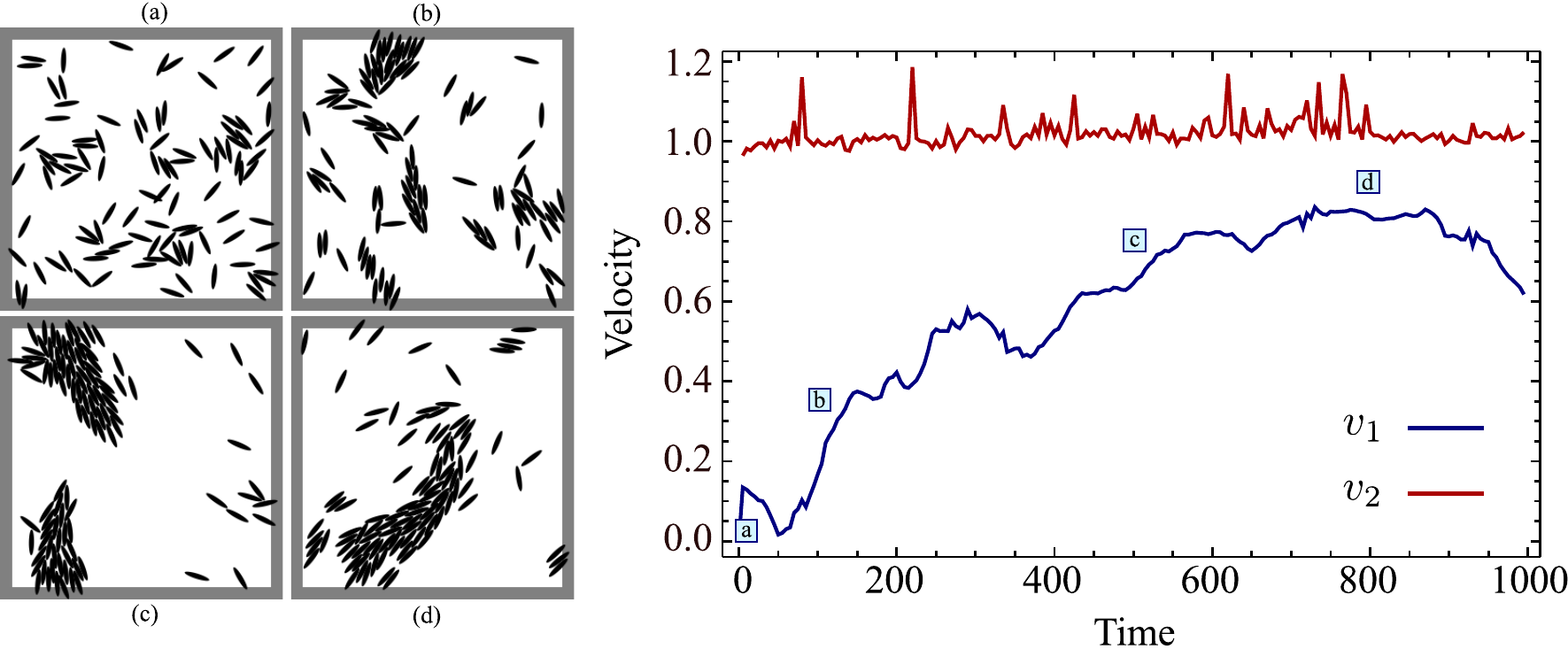}	
\caption{\label{fig:periodic-boundary}Example of {\em flocking} in a group of 100 ellipses from a numerical solution of Eqs. \eqref{eq:model} on a periodic square domain of size $L/a=100$. The right panel shows the evolution of the order parameter $v_{1}$ and $v_{2}$ as a function of time. The parameter values used are: $\omega=0$, $v_{0}=1$, $k_{1}=10$, $k_{2}=0.2$, $\zeta=1$.}
\end{figure}

Varying the two relevant parameters in Eq. \eqref{eq:model} and integrating them numerically leads to a variety of collective behaviors consistent with our observations as shown in Fig. \ref{fig:collective-behavior-th} (\ref{sec:appendix-b} and Supplementary Movie 5). We see that individual self-propelled walkers or spinners tend to migrate toward the boundary of the arena where they experience a torque that reorients the individual toward the interior. At low densities, the primary interactions of these automata are with the boundary and so one sees random uncoordinated movements (Fig. \ref{fig:collective-behavior-th}a). At higher densities, as more ellipses simultaneously cluster in the same region of the boundary, the aligning force exerted on the ellipses by each other can overcome the action of the boundary provided the cluster is large enough. Thus walkers for whom $\omega a/v_{0}\sim 0$ tend to aggregate into a static structure at the boundary at high enough density (Fig. \ref{fig:collective-behavior-th}c).  However, spinners for whom $\omega a/v_{0} > 0$ form clusters at the boundary that are tilted, and this broken symmetry together with the effect of the weak topography (boundary curvature) keeps them confined to the neighborhood of the boundary and causes the automata to eventually synchronize their velocities resulting in a collective swirling motion of the entire cluster (Fig. \ref{fig:collective-behavior-th}b). \revision{In absence of confinement, our system of self-propelled ellipses shows the typical flocking behavior of the Vicsek model (Vicsek {\em et al}. 1995), so that for large density and small noise, the particles organizes in lanes or coherently moving subunits (see Fig. \ref{fig:periodic-boundary} and Supplementary Movie 6). We note that both collective swirling stasis originates from the interplay between self-propulsion, particle geometry and confinement and do not occur in systems without boundary.} 	

Our model described by Eq. \eqref{eq:model} illustrates the origin of the three observed behaviors in a broader context. Analogously to the simple self-propelled particles, BBots tend to migrate to the boundary, which depending on the local density of the BBots and their angular velocity, can either play the role of an obstacle that causes the objects to jam, or a confining channel that collects and aligns the BBots into a coordinated moving cluster. For a cluster of walkers at the boundary, each BBot in the cluster is trapped by its neighbors and cannot escape. As their angular velocity increases, they can exert a sufficient torque on their neighbors to push them aside and escape from the cluster, consistent with the observation that the number of BBots required for jamming decreases with their angular velocity, as shown in the phase diagram in Fig. \ref{fig:phase-diagram}. Bbots that are spinners have a relatively small translational velocity and so are easily trapped by their neighbors at the boundary once their density is large enough and they unable to reverse direction and escape. However, the finite spinning torque leads to a global tilt of the BBots leading to a global swirling motion of the entire cluster along the edge of the arena.

\section{Discussion}

While the geometric structure of the clusters depends significantly on the shape of the particles, the occurrence of these three collective behaviors observed in the experiment is rather general. To demonstrate this we have run an additional set of simulations using self-propelled polar disks in place of elliptical particles (see \ref{sec:appendix-b}). The disks' dynamics is also dictated by Eq. \eqref{eq:model}, in which the physical torque is $M_{ij}$ now replaced by a generic aligning interaction of the form: $M_{ij}=\sin(\theta_{j}-\theta_{i})/N_{i}$ (which has no real physical basis in our system). With this choice, Eq. (\ref{eq:model}b) becomes a short-range version of the Kuramoto model for phase-synchronization in chemical and biological oscillators (Acebr\'on {\em at} 2005). Analogously to the ellipses, the dynamics of the polar disks is also characterized by three regimes: random motion at low densities, jamming at the boundary for large densities when the angular velocity of the disks vanishes, and the formation of a compact cluster that circulates along the boundary for large densities when the disks have a finite angular velocity (see \ref{sec:appendix-b}).

The similarity between our experimental system and the two models described above suggests that the coordinated circulation and jamming in a system of confined agents is generic. This form of collective behavior relies on simple but crucial features of the individual agents as well as the environment: the ability to translate and rotate, and the ability to interact with each other and with the environment, here including the boundary and the local topography. While the spatial structure of the clusters crucially depends on the shape and the packing properties of the particles, their collective motion is very robust, and depends on simple non-specific principles. 

In living systems, where similar behaviors such as the density-driven transitions are seen in confined {\it Macrotermes michaelseni} (Turner 2011), they have been linked with insect cognition and social interactions. Our study suggests that particle motion, shape and spatial interactions are sufficient and might in fact play equivalent roles. \revision {In a biological setting such as termite swarms, one might test these ideas by controlling the confinement of termites by varying the substrate curvature and slipperiness, gluing circular discs on their backs to make the interactions more isotropic, etc.}  In an artificial setting, the collective abilities of spinner and walker BBots to convert environmental interactions into dynamical behavior may allow us to explore functional swarms that can search and sense environments. For example,  they have the ability to sense substrate roughness by slowing down, and they can  search topography (curvature) in massively parallel ways, using mechanical intelligence, and suggesting the use of these automata as fast, cheap, leaderless explorers.
 
\section*{Acknowledgments}

We thank the Wyss Institute, the Harvard -Kavli Nano-Bio Science and Technology Center, the Harvard-NSF MRSEC, DARPA, VARTA Microbattery and the MacArthur Foundation for support, J. McArthur for constructing the charger used to recharge the batteries of the BBots,  Teo Guo Xuan for Fig. 1 (left top), Joe Ustinowich and Anas Chalah for invaluable support in fabricating our bots, and A. Mukherjee and M. Bandi for many useful discussions and suggestions.
	
\appendix{\label{sec:appendix-a}Locomotion of an individual BBot}	
	
The principle of motion of a single BBot, as inferred from high speed videos, relies on the sequence of events illustrated in Fig. \ref{fig:single-bot}. At each cycle of the eccentric motor the following sequence of events takes place: 1) the bristles bend as they are loaded by a force $F=Mg+mr\omega^2$, where $Mg$ is the weight of the BBot, $m$ the eccentric mass, $r$ the lever arm and $\omega$ the angular frequency of the motor. 2) While the eccentric mass rotates, the load on the bristles decreases; this causes the bristles to unbend. 3) The unbending bristles slip on the underlying substrate, producing a forward displacement of the object in the horizontal direction. 

A quantitative description of the gait reduces to calculating the horizontal displacement $\Delta x$ of the bristles at each cycle of the eccentric motor. To accomplish this we ignore the collective dynamics of the bristles and focus on a planar description, replacing the rows of bristles as an ideal elastic rod subject to periodic tip-load acting in the $y$ direction:
\begin{equation}\label{eq:load}
F_{y} = W = Mg+mr\omega^{2}\sin \omega t\,. 
\end{equation}
When the eccentric mass is oriented with its axis of symmetry toward the negative $y$-direction, the load is $W_{\max}=Mg+mr\omega^{2}$ and the bristles are maximally deflected. As the eccentric mass moves away from the vertical direction, the bristles start to recoil and their tip slides on the substrate. The sliding tip of the bristles is subject to a dynamic frictional force acting along the $x$-direction:
\begin{equation}\label{eq:friction}
F_{x} = \mu W\,,
\end{equation}
where $\mu$ is kinetic friction coefficient. The unbending of the bristles terminates when the eccentric mass is oriented along the positive $y$-direction and the load is minimal: $W_{\min}=Mg-mr\omega^{2}$. 

To make progress we assume that the inertia of the bristles is negligible so that bristle deflection and sliding occurs {\em quasistatically}. This implies that the conformation of the bristles is, at any time of the gait cycle, in equilibrium with the external load and the frictional force acting respectively on the $y$ and $x$ direction. Under this assumption, the shape of the bristles is governed by the classical equilibrium equations of an ideal elastic beam:
\begin{equation}\label{eq:euilibrium}
\bm{F}_{s}+\bm{K} = 0\,,\qquad
\bm{M}_{s}+\bm{t}\times\bm{F} = 0\,.	
\end{equation}
$\bm{F}$ and $\bm{M}$ are respectively the force and torque per unit length and the subindices denote a derivative with respect to the arc-length $s$ of the beam. The tangent vector $\bm{t}$ of the bristles is given by:
\begin{equation}\label{eq:tangent-vector}
\bm{t}= \sin\theta\,\bm{\hat{x}}-\cos\theta\,\bm{\hat{y}}\,,
\end{equation}
where $\theta$ is the angle formed by the bristles with the vertical direction. Finally, $\bm{K}$ is the external force acting on the tip of the bristles, thus:
\begin{equation}\label{eq:external-force}
\bm{K} = W(\mu\,\bm{\hat{x}}+\bm{\hat{y}})\delta(s-L)\,,
\end{equation}%
\begin{figure}[t]
\centering
\includegraphics[width=0.5\textwidth]{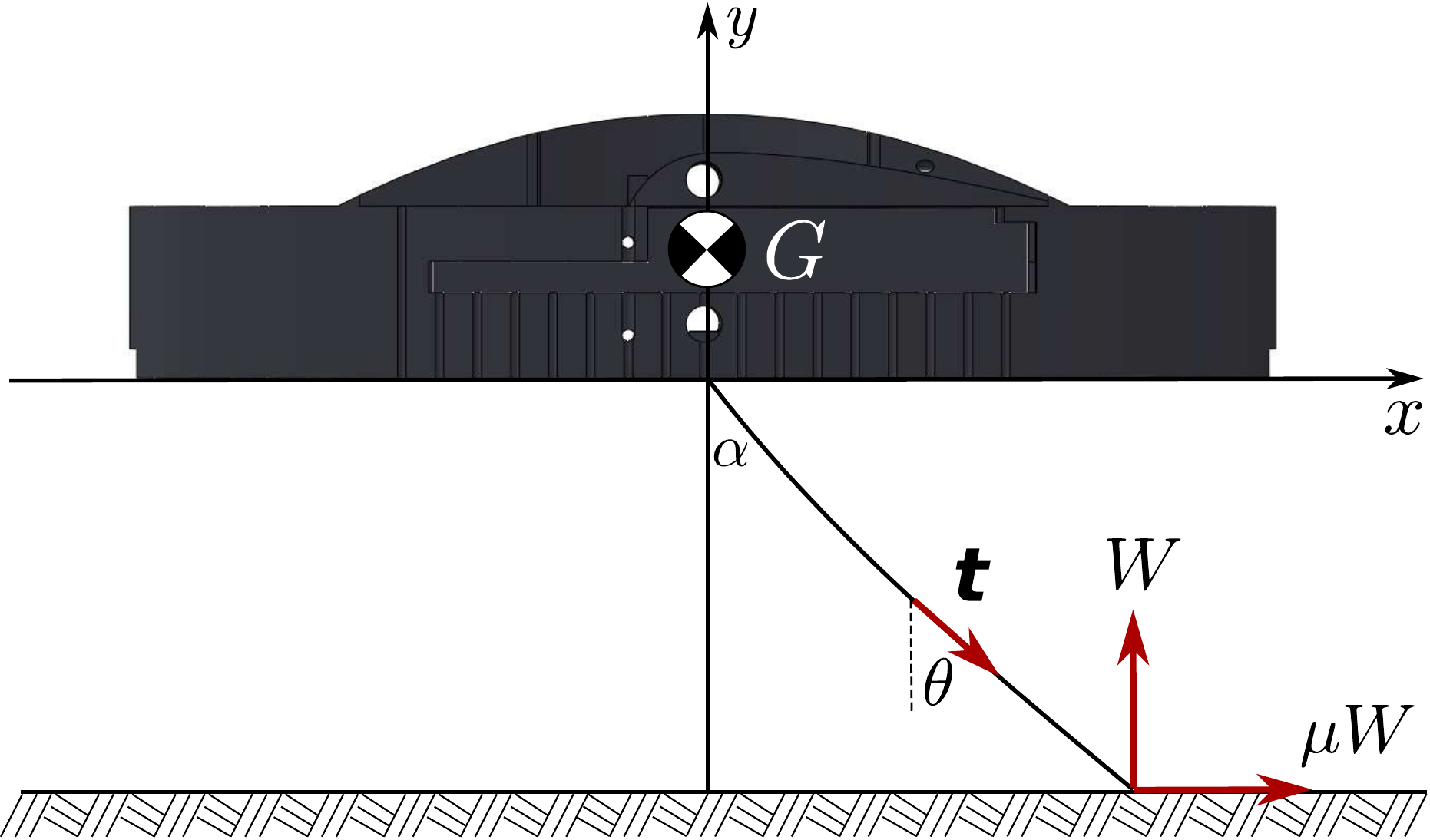}	
\caption{\label{fig:bbot}Schematic representation of BBot principle of motion. The two rows of bristles are modeled as a single elastic beam whose free end is subject to two forces: the time dependent weight $W=Mg+mr\omega^{2}\sin\omega t$ acting on the positive $y$-direction and a kinetic frictional force acting on the positive $x$-direction. The gait cycle is assumed {\em quasistatic} so that the shape of the bristle is, at any time, in equilibrium with the applied forces.}
\end{figure}%
where $L$ is the length of the bristles and the delta function reflects the fact that the force is applied at the tip. Integrating the force equation and replacing it in the torque equation gives:
\begin{equation}\label{eq:torque}
\bm{M}_{s} = W\sin\theta\,\bm{\hat{z}}+\mu W \cos\theta\,\bm{\hat{z}}\,.
\end{equation}
The torque $\bm{M}$ acting in the beam and its curvature $\kappa=\theta_{s}$ are related by the Euler-Bernoulli constitutive equation $\bm{M}/EI=-\kappa\bm{b}$, where $\bm{b}$ is the binormal vector of the beam ($\bm{\hat{z}}$ in this case) and $EI$ is its bending rigidity (with $E$ the Young modulus and $I$ the area-moment of inertia). This yields a single differential equation for the angle $\theta$:
\begin{equation}\label{eq:theta-eq1}
EI\,\theta_{ss}+W\sin\theta+\mu W\cos\theta = 0\,,
\end{equation}
with boundary conditions:
\begin{equation}\label{eq:boundary-condition}
\theta(0) = \alpha\,,\qquad
\theta_{s}(L) = 0\,.
\end{equation}
These are the typical boundary conditions of a cantilever beam, with one end fixed at an angle $\alpha$ and the other free of torques. It is convenient to work with dimensionless quantities, by rescaling the arc-length with the total length of the bristles: $t=s/L$, so that Eq. \eqref{eq:theta-eq1} can then be recast in the form
\begin{equation}\label{eq:theta-eq2}
\theta_{tt} + k^{2}\sin(\theta+\varphi) = 0\,,	
\end{equation}	
where:
\begin{equation}\label{eq:definitions}
k^{2} = \frac{W}{EI/L^{2}}\,\sqrt{1+\nu^{2}}\,,\qquad
\varphi = \arctan \mu\,.
\end{equation}
Eq. \eqref{eq:theta-eq2} can be integrated exactly in terms of Jacobi elliptic functions to yield
\begin{subequations}\label{eq:solution}
\begin{align}
&\sin \tfrac{1}{2}(\theta+\varphi) = m \sn\left(k(t-1)+K,m\right)\,, \\[10pt]
&\cos \tfrac{1}{2}(\theta+\varphi) = \dn\left(k(t-1)+K,m\right)\,, \\[10pt]
&\theta_{t} = 2mk \cn\left(k(t-1)+K,m\right)\,,
\end{align}	
\end{subequations}
where Eqs. \eqref{eq:solution} follow using standard techniques (see for instance Davis 1960) and we use the standard notation for elliptic functions and integral, i.e. given the elliptic integral of the first kind:
\begin{equation}
u = F(\phi,m)= \int_{0}^{\phi} \frac{dt}{\sqrt{1-m^{2}\sin^{2}t}}\,,
\end{equation}
with $0<m^{2}<1$ the elliptic modulus and $\phi$ is the Jacobi amplitude: $\phi = \am(u,m)$. From this it follows that
\begin{equation}
\sn(u,m) = \sin\phi\,, \qquad		
\cn(u,m) = \cos\phi\,, \qquad
\dn(u,m) = \sqrt{1-m^{2}\sin^{2}\phi}\,. 
\end{equation}	
\begin{figure}[t]
\centering
\includegraphics[width=1\textwidth]{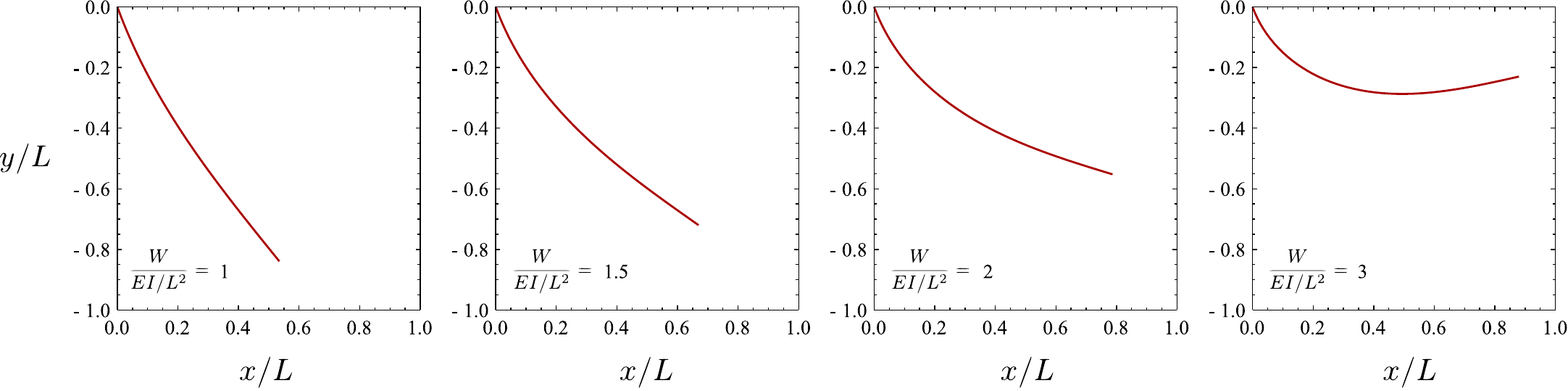}
\caption{\label{fig:elastica}Example of configurations of the deflected bristles obtained from Eqs. \eqref{eq:xy} for various $\frac{W}{EI/L^{2}}$ values. The orientation of the bristles at the supported end is $\alpha=20^{\circ}$ and the friction coefficient is $\mu=0.1$.}	
\end{figure}
Finally, the quantity $K$ in Eqs. \eqref{eq:solution}, is the complete elliptic integral of the first kind: $K = F(\frac{\pi}{2},m)$. This enforces the boundary condition at the free end $t=1$:
\begin{equation}
\theta_{t}(1) = 2mk \cn\left(K,m\right) = 0\,.
\end{equation}
The elliptic modulus $m$, on the other hand, is obtained from the boundary condition at the fixed end through Eq. (\ref{eq:solution}a):
\begin{equation}\label{eq:elliptic-modulus}
\sin\tfrac{1}{2}(\alpha+\varphi) = m \sn(K-k,m)\,.
\end{equation}
With the solution \eqref{eq:solution} in hand, we can now construct a parametric equation for the shape of the deflected bristles by integrating the tangent vector $\bm{t}$:
\begin{equation}\label{eq:position-vector}
\bm{r}(s') = \int_{0}^{s'}ds\,\bm{t}(s) = \int_{0}^{s'} ds\,(\sin\theta\,\bm{\hat{x}}-\cos\theta\,\bm{\hat{y}})\,. 
\end{equation}
In order to use Eq. \eqref{eq:position-vector} we first set $A=\sin\frac{1}{2}(\theta+\varphi)$ and $B=\cos\frac{1}{2}(\theta+\varphi)$ and note that:
\begin{subequations}\label{eq:AB}
\begin{align}
\cos\theta &= 2AB\sin\varphi+(B^{2}-A^{2})\cos\varphi\,, \\[10pt]
\sin\theta &= 2AB\cos\varphi-(B^{2}-A^{2})\sin\varphi\,, 
\end{align}
\end{subequations}
and that the integrals of terms containing $A$ and $B$ are given, up to a constant, by:
\begin{subequations}\label{eq:integrals}
\begin{gather}
\int dt\,AB = -\frac{m}{k}\cn\left(k(t-1)+K,m\right)\,, \\[10pt]
\int dt\,(B^{2}-A^{2}) = -t + \frac{2}{k}\,E\left(\am\left(k(t-1)+K,m\right),m\right)\,,
\end{gather}
\end{subequations}
where $E$ is the elliptic integral of the second kind, defined as:
\begin{equation}
E\left(\phi,m\right) = \int_{0}^{\phi}dt\,\sqrt{1-m^{2}\sin^{2}t}\,.	
\end{equation}
Then, combining Eqs. \eqref{eq:AB} and \eqref{eq:integrals} we obtain the following parametric expression of the coordinate of the bristles:
\begin{subequations}\label{eq:xy}
\begin{align}
x(t)/L &= x_{0}/L-\frac{2m}{k}\,\cn\left(\tau,m\right)\cos\varphi+\left[t-\frac{2}{k}\,E\left(\am\left(\tau,m\right),m\right)\right]\sin\varphi\,,\\[10pt]
y(t)/L &= y_{0}/L+\frac{2m}{k}\,\cn\left(\tau,m\right)\sin\varphi+\left[t-\frac{2}{k}\,E\left(\am\left(\tau,m\right),m\right)\right]\cos\varphi\,,
\end{align}
\end{subequations}
where we have called $\tau=k(t-1)+K$ for brevity. The integration constants $x_{0}$ and $y_{0}$ are set so that $x(0)=y(0)=0$ so that
\begin{subequations}\label{eq:constant}
\begin{align}
x_{0}/L=\frac{2}{k}\,E\left(\am\left(K-k,m\right),m\right)\sin\varphi+\frac{2m}{k}\,\cn\left(K-k,m\right)\cos\varphi\,,\\[10pt]
y_{0}/L=\frac{2}{k}\,E\left(\am\left(K-k,m\right),m\right)\cos\varphi-\frac{2m}{k}\,\cn\left(K-k,m\right)\sin\varphi\,,
\end{align}
\end{subequations}
Eqs. \eqref{eq:xy}-\eqref{eq:constant}, and \eqref{eq:elliptic-modulus} along with the definitions \eqref{eq:definitions} give the shape of the bristles. In Fig. \ref{fig:elastica} we show a sequence of typical configurations obtained from this solution for various values of $\frac{W}{EI/L^{2}}$.

\begin{figure}[t]
\centering
\includegraphics[width=0.9\textwidth]{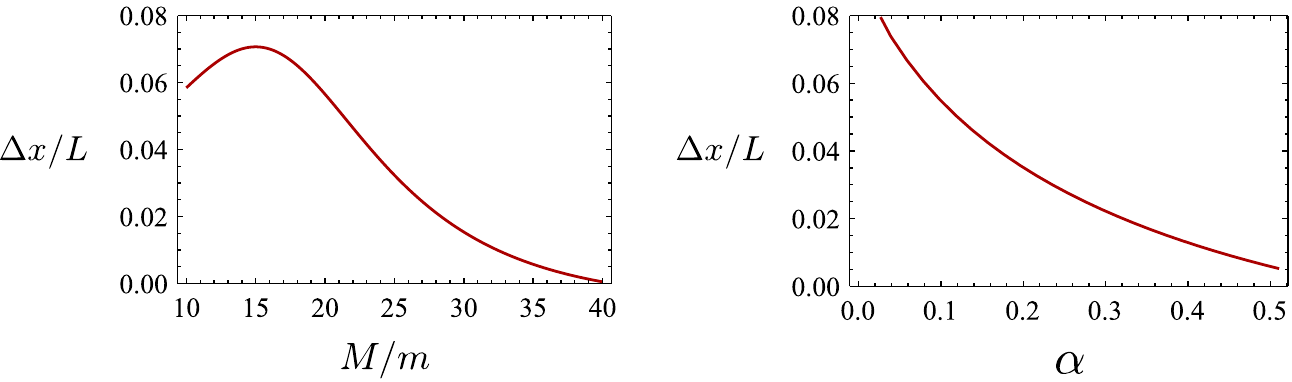}	
\caption{\label{fig:stepsize}The step size $\Delta x/L$ as a function of the ratio $M/m$ (left) and the bristle angle $\alpha$ from Eqs. \eqref{eq:xy}. The parameters used for the plots are $g/r\omega^{2}=1$ and $\frac{mr\omega^{2}}{EI/L^{2}}=0.1$.}
\end{figure}

Given the shape of the bristles, the step size $\Delta x$ of a Bbot associated with each gait cycle is dictated by the position of the tip of the bristles. The latter can be obtained from Eqs. \eqref{eq:xy} by setting $t=1$ and noting that $\cn(K,m)=0$, $\am(K,m)=\pi/2$. Then
\begin{equation}
x(1)= x_{0}+L\sin\varphi\left[1-\frac{2E(m)}{k}\right]\,,
\end{equation}
where $E(m)=E(\frac{\pi}{2},m)$ is the complete elliptic integral of the second kind. Because of our quasistatic approximation, $x(1)$ depends exclusively on the applied load and so the step size is simply given by the difference in the position of the tip associated with the maximal (eccentric motor in the negative $y$-direction) and minimal (eccentric motor in the positive $y$-direction) load. Using the definition
\begin{equation}
k^{2}_{\pm} = \frac{Mg \pm mr\omega^{2}}{EI/L^{2}}\,\sqrt{1+\mu^{2}}\,,		
\end{equation}
we can finally express the step size in the form:
\begin{equation}\label{eq:stepsize}
\Delta x = \Delta_{0}-2L\sin(\arctan\mu)\left[\frac{E(m_{+})}{k_{+}}-\frac{E(m_{-})}{k_{-}}\right]\,,
\end{equation}
where $m_{\pm}=m(k_{\pm})$ and $\Delta_{0}=x_{0}(k_{+})-x_{0}(k_{-})$. The model is valid only as long as $k_{-}^{2}>0$, which implies $Mg>mg\omega^{2}$. 
For light BBots, where this condition does not hold,  locomotion is complicated by the fact that when the eccentric mass is oriented along the positive $y$-direction, there is an upward directed force that makes the BBot lose contact with the substrate. The resulting jumping motion then couples with the dynamics of the bristles making the gait cycle intractable with the methods used here. Fig. \ref{fig:stepsize} shows a typical step size obtained from Eqs. \eqref{eq:xy} as a function of the ratio $M/m$ and the bristles inclination angle $\alpha$. 

Our analytical results allow us to capture the qualitative aspects of the motion of a single Bbot and its dependence on the magnitude and frequency of the eccentric driving motor, as well as the dependence on the mass of the Bbot, the orientation and length of the bristles, consistent with experimental observations. An alternative analysis of the locomotion of an individual BBot was carried out by DeSimone \& Tatone (2012) using methods of geometric control theory (see Alouges {\em et al}. 2008).

\appendix{\label{sec:appendix-b}Collective behavior of self-propelled disks}

In order to gain insight into the origin and the generality of the behaviors observed in our experiments and numerical simulations of interacting Bristle-Bots (BBots), we also compared the results with those obtained from the numerical simulation of self-propelled disk-like particles that are isotropic. The particles have both a positional degree of freedom given by their center of mass $\bm{r}_{i}$ and an orientation $\bm{n}_{i}=(\cos\theta_{i},\sin\theta_{i})$ with the position $\bm{r}_{i}$ and the angle $\theta_{i}$ that evolve according to Eqs. \eqref{eq:model}, but, in contrast with the case of elliptical collisions where there is a physical torque that causes alignment, $M_{ij}$ is chosen to be:
\begin{equation}
M_{ij} = \sin(\theta_{j}-\theta_{i})/N_{i}\,.	
\end{equation}
With this choice, Eq. (\ref{eq:model}b), is a short-range version of the Kuramoto model for phase-synchronization (Acebr\'on {\em et al}, 2005) and can serve as a rather general model for aligning interactions among self-propelled particles, although it has no direct physical origin. 

\begin{figure}[t]
\centering
\includegraphics[width=0.9\textwidth]{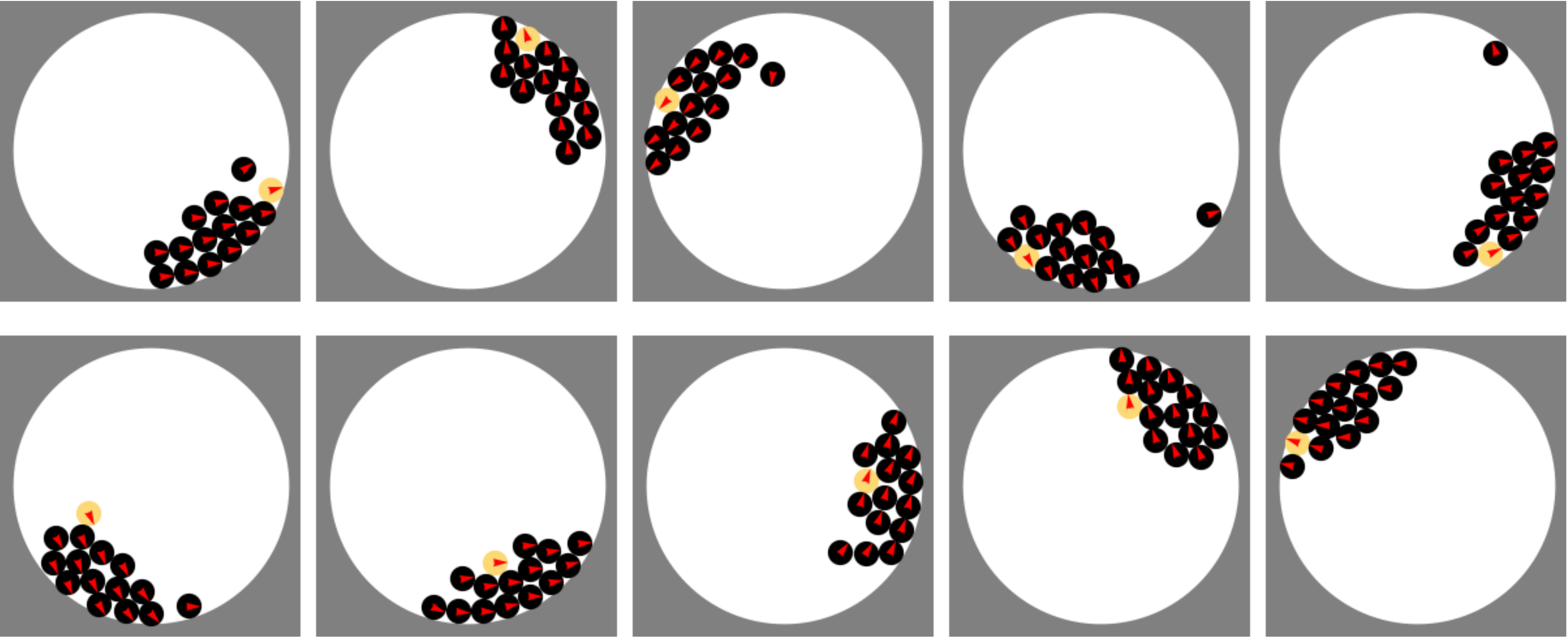}	
\caption{\label{fig:treadmilling}Example of treadmilling in a group of 15 polar disks obtained from a numerical solution of Eqs. \eqref{eq:model}. The top frames shows the regression of a {\em tracer} disk (labeled in yellow) from the leading to the trailing end of the cluster. The {\em tracer} disk remains in proximity of the trailing end for about a loop and then starts its progression toward the front along internal side of the cluster. The parameter values used are: $\omega=0.05$, $v_{0}=1$, $k_{1}=10$, $k_{2}=3$, $k_{3}=0.1$, $a=1$, $R=11$, $\xi=2a$, $\zeta=5\pi/2$.}
\end{figure}

For our simulations, we assume that the particles are confined to a circular domain of radius $R$ centered at the origin. The interaction between the particles and the boundary that takes place is assumed to have an identical form to that used in the simulations of the elliptical particles.

\begin{figure}[t]
\centering
\includegraphics[width=0.9\textwidth]{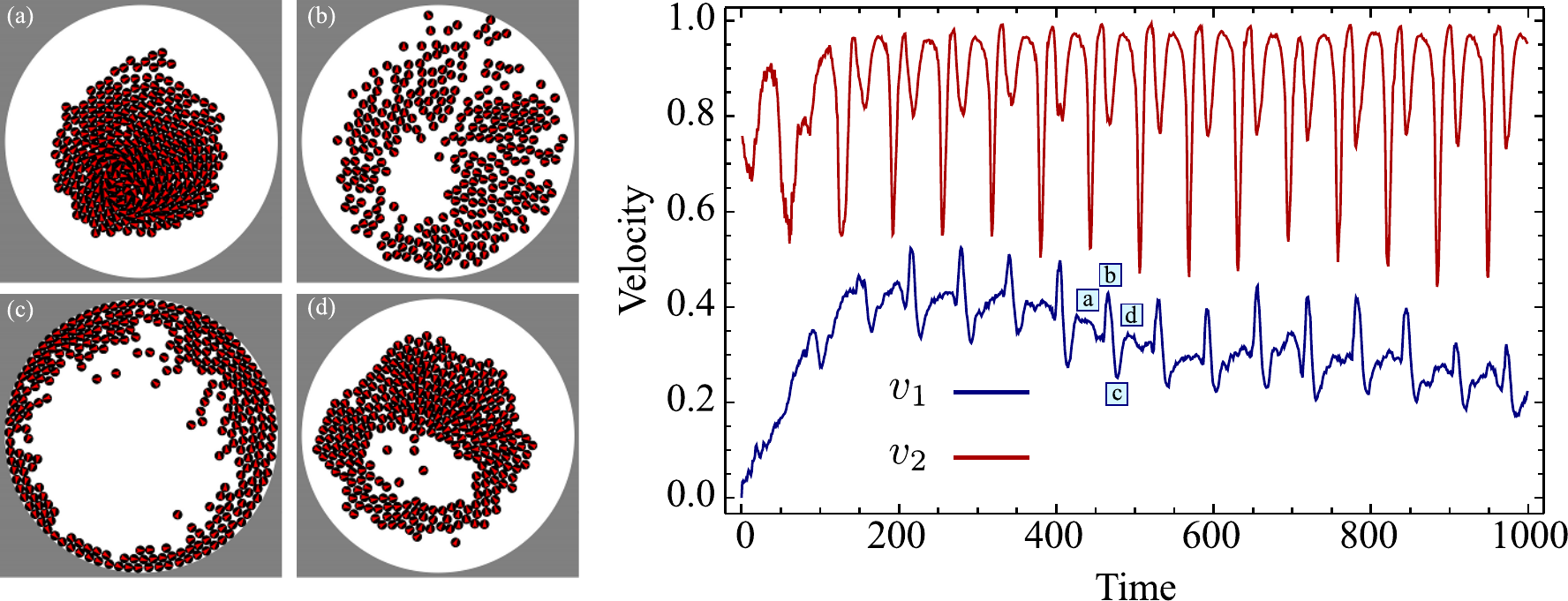}	
\caption{\label{fig:breathing}Example of {\em breathing} in a group of 300 polar disks obtained from a numerical solution of Eqs. \eqref{eq:model}. The right panel shows the evolution of the order parameter $v_{1}$ and $v_{2}$ as a function of time. The dips in the $v_{2}$ trajectory correspond to the configurations where the disks are densely packed at the center of the arena (pannel {\em a} in the left side of the figure), while the peaks in the $v_{1}$ trajectory denote the bursts as a consequence of which the disks suddenly migrate to the boundary. The parameter values used are: $\omega=0.1$, $v_{0}=1$, $k_{1}=10$, $k_{2}=0.2$, $k_{3}=0.1$, $a=1$, $R=28$, $\xi=a$, $\zeta=\pi$.}
\end{figure}

While the spatial structure and packing properties of the clusters depends on the details of the system, and in particular on the shape of the particles, the occurrence of the coordinated behaviors (swarming, swirling and stasis) appears to be a very robust property of systems of self-propelled agents in a confined space. These behaviors are not sensitive to the presence of inertia [this is present in the experiment and is neglected in Eqs. \eqref{eq:model}] or to the shape of the particles and the precise form of the aligning torque $M_{ij}$. However, there are features that do depend on the details; a most interesting example is what we term {\em treadmilling}, observed in the numerical simulation of both ellipses and disks. Fig. \ref{fig:treadmilling} shows an example of treadmilling in a group of polar disks, wherein particles move through the cluster and eventually leave it, only to join it later at the other end. For very large densities, the self-propelled disks also exhibit a {\em breathing mode}, in which the particles periodically migrate from the center to the boundary and vice-versa by mean of sudden bursts (Fig. \ref{fig:breathing}) reminiscent of those observed in {\em excitable} active systems (see Giomi {\em et al}. 2011 and 2012).

\end{document}